\documentstyle[12pt]{article}
\def\lsim{\mathrel{\lower2.5pt\vbox{\lineskip=0pt\baselineskip=0pt
           \hbox{$<$}\hbox{$\sim$}}}}
\def\gsim{\mathrel{\lower2.5pt\vbox{\lineskip=0pt\baselineskip=0pt
           \hbox{$>$}\hbox{$\sim$}}}}
\begin{document}
\setlength{\baselineskip}{8mm}
\begin{titlepage}
\begin{flushright}
\begin{tabular}{c c}
& {\normalsize  DPNU-96-49} \\
& {\normalsize September 1996}
\end{tabular}
\end{flushright}
\vspace{5mm}
\begin{center}
{\large \bf Can the Higgs Sector Trigger $CP$ Violation in the MSSM?}\\
\vspace{15mm} 
Naoyuki Haba\footnote{E-mail:\ haba@eken.phys.nagoya-u.ac.jp} \\ 

{\it 
Department of Physics, Nagoya University \\
           Nagoya, JAPAN 464-01 \\
}
\end{center}

\vspace{10mm}


\begin{abstract}

We reanalyze the possibility of 
$CP$ violation in the Higgs sector of the 
minimal supersymmetric standard model (MSSM). 
Contrary to the result of previous analysis, 
spontaneous $CP$ violation can not occur 
by only chargino and neutralino 
radiative corrections since 
the vacuum does not stable. 
Top and stop radiative corrections are crucially needed. 
However even with this correction 
there is no experimentally allowed region 
in $\tan \beta \geq 1$. 
This situation is not remedied even if 
the stop left-right mixing is included. 
We also analyze explicit $CP$ violation in the 
Higgs sector of the MSSM and show that the 
effect is too small 
to influence the phenomenology. 
We thus show that the Higgs sector can not, 
by itself, trigger $CP$ violation in the MSSM. 

\end{abstract}
\end{titlepage}

%
\section{Introduction}

The minimal supersymmetric standard model (MSSM) 
is one of the strongest candidate for physics beyond the 
standard model (SM). 
Since the MSSM naturally contains two Higgs doublets, 
$CP$ could be violated in the Higgs sector 
both spontaneously and explicitly. 
The possibility of spontaneous $CP$ violation in the MSSM, 
which is caused by the 
non-trivial phase of vacuum expectation values (VEVs), 
was first discussed in Refs.\cite{MAEKAWA}\cite{POMAROL}. 
They have shown the following; \\
(i) Spontaneous $CP$ violation is caused essentially 
by the chargino and the neutralino 
radiative corrections. \\
(ii) This scenario is excluded from 
the experiment since the "pseudoscalar" mass 
is of $O(\sqrt{\lambda_5} v)$ which is about 5 GeV. 
\par
In this paper we reconsider (i) and (ii), 
and results are the following;  \\
\par
\noindent 
{\bf (a)} $CP$ violating vacuum discussed in 
Ref.\cite{MAEKAWA} and \cite{POMAROL} is not stable. 
It becomes stable when 
top and stop contributions are added, 
provided that the stop mass is larger than 
180 GeV in the limit of small stop 
left-right mixing. \\
\par
\noindent
{\bf (b)} Result (ii) was obtained in the case where 
there is no left-right mixing of stop. 
If the effect of the stop left-right mixing is included, 
Higgs masses depend on 
new parameters $A_t$ and $\mu$. 
Thus, there might be the experimental 
allowed region for spontaneous $CP$ violation scenario 
in the MSSM. 
Unfortunately, however, 
the numerical analysis shows that 
$CP$ violating vacuum is 
unstable in the parameter region of 
$A_t /m_{\tilde t}$ and/or 
$\mu /m_{\tilde t} \geq O(1/3)$, 
and the situation is almost same as 
the small left-right mixing limit case 
when $A_t /m_{\tilde t}$ and 
$\mu /m_{\tilde t} \leq O(1/3)$. 
We find that the result (ii) does not change 
even when we consider the possibility 
of left-right mixing. \\
\par
\noindent
{\bf (c)} In this paper 
we consider only the region 
$\tan \beta \geq 1$
\footnote{Parameter space $\tan \beta < 1$ is strongly 
disfavored in low energy SUSY 
models\cite{NOJIRI} as pointed out in Ref.\cite{POMAROL}.}. 
As for the experimental constraints, 
since spontaneous $CP$ violation 
makes scalars and a pseudoscalar mix, 
it is not accurate to compare the lightest Higgs mass 
predicted in this scenario to the 
lower limit of pseudoscalar mass 24.3 GeV in 
Ref.\cite{PDG} in which they assume that 
scalar and pseudoscalar do not mix. 
Therefore, 
we need to consider 
the precise experimental constraints from 
(A): $Z \rightarrow h_1h_2$ and 
(B): $Z \rightarrow h_i l^+l^-$ $(i=1,2)$. \\
\par
\noindent
{\bf (d)} We also discuss briefly 
explicit $CP$ violation 
in the Higgs sector of the MSSM 
in this paper. 
Since it 
is also caused by the radiative correction, 
its effect is too small to 
influence the phenomenology. \\
\par
We therefore conclude that this scenario 
is excluded and the Higgs sector can not, 
by itself, trigger $CP$ violation. 
\par
In Section 2, we discuss spontaneous $CP$ violation. 
Section 3 gives summary and discussion. 
In Appendix, we show explicit $CP$ violation in the Higgs sector. 
%
%
\section{Reanalysis of Spontaneous $CP$ Violation in the MSSM}
The most general 
two Higgs doublet model potential\cite{THDM} is given by 
\begin{eqnarray}
\label{TDLEE}
 V(H_1, H_2) &=& m_1^2 |H_1|^2 + m_2^2 |H_2|^2 -
        (m_{12}^2 H_1 H_2 + {\rm h.c.})         
       + \lambda_1 |H_1|^4 + \lambda_2 |H_2|^4   \nonumber  \\
   &+&   \lambda_3 |H_1|^2|H_2|^2              
       + \lambda_4 |H_1 H_2|^2 +
        {1\over2}[\lambda_5 (H_1 H_2)^2 + {\rm h.c.}]   \\ 
   &+& {1\over2}[\lambda_6 (H_1 H_2)|H_1|^2 + {\rm h.c.}] +
       {1\over2}[\lambda_7 (H_1 H_2)|H_1|^2 + {\rm h.c.}] .
                                                       \nonumber  
\end{eqnarray}
$H_1$ and $H_2$ are Higgs doublet fields denoted as 
\begin{equation}
H_1=\left(
\begin{array}{c}
H_1^{0} \\
H_1^- \\
\end{array}
\right) , \qquad
H_2=\left(
\begin{array}{c}
H_2^+ \\
H_2^0 \\
\end{array}
\right),
\end{equation}
with 
\begin{equation}
  H_1H_2=H_1^0H_2^0-H_1^-H_2^+. 
\end{equation}
Quartic couplings $\lambda_i$s $(i = 1 \sim 4)$ are 
written by gauge couplings in the MSSM as 
\begin{equation}
\label{lambda1234}
\lambda_1 = \lambda_2 = {1\over8}(g^2+g'^2), \;\;
\lambda_3 = {1\over4}(g^2-g'^2), \;\; 
\lambda_4 = -{1\over2}g^2 ,  
\end{equation}
where $g$ and $g'$ are gauge couplings of $SU(2)_L$ and $U(1)_Y$, 
respectively. 
Parameters $m_1$, $m_2$, and $m_{12}$ are arbitrary 
determined by supersymmetric Higgs mass $\mu$ 
and soft SUSY breaking parameters. 
Coupling $\lambda_{5}$ get non-zero positive 
value by radiative corrections of 
the chargino and the neutralino\cite{MAEKAWA}\cite{POMAROL}. 
The value of $\lambda_5$ is 
\begin{equation}
\label{lambda5}
\lambda_5 = {g^4 \over 32 \pi^2} \sim 5 \times 10^{-4}, 
\end{equation}
in the limit of small squark left-right mixings and 
SUSY breaking mass parameter $B$, 
and equal mass limit of charginos 
and neutralinos\cite{POMAROL}. 
Couplings $\lambda_{6}$ and $\lambda_{7}$ are expected to 
be the same order of $\lambda_5$. 
Parameters $m_{12}^2$ and $\lambda_{5 \sim 7}$ 
are all complex in general. 
In the case that the Higgs sector has $CP$ symmetry, 
\begin{equation}
\label{lambdarelation}
{\rm Im}(\lambda_5^* m_{12}^4) = 
{\rm Im}(\lambda_5^* \lambda_6^2) = 
{\rm Im}(\lambda_5^* \lambda_7^2) =0  
\end{equation}
are satisfied, and 
all these parameters can be real by 
the redefinition of Higgs fields. 
Then we can set all parameters to be real 
in spontaneous $CP$ violation scenario. 
As for the explicit $CP$ violation, 
Eq.(\ref{lambdarelation}) is not held as 
shown in Appendix. 
Eq.(\ref{lambdarelation}) shows that the 
Higgs potential of the MSSM is automatically 
$CP$ invariant in the tree level 
because $\lambda_{5 \sim 7}^{\rm (tree)}=0$ . 
\par
Assuming that the charged Higgs 
does not get VEV, 
we denote VEVs of neutral components as 
\begin{equation}
\label{VEVs}
 \langle H_1^0 \rangle = v_1 , \;\;\;\;\; 
 \langle H_2^0 \rangle = v_2 e^{i \phi} ,
\end{equation}
where $v_1$ and $v_2$ are 
real and positive parameters which satisfy 
$v \equiv \sqrt{v_1^2+v_2^2}=174$ GeV. 
We define fields around this vacuum as 
\begin{eqnarray}
 \label{fielddifinition0}
  H_1^0 &=& v_1 + 
       {1 \over \sqrt{2}} (S_1+i \sin \beta A), 
                                       \nonumber \\
 \label{fielddifinition1}
  H_2^0 &=& v_2 e^{i \phi} + 
       {1 \over \sqrt{2}}e^{i \phi} (S_2+i \cos \beta A), 
\end{eqnarray}
where $S_1$ and $S_2$ are scalar fields 
and $A$ is a pseudoscalar field, and 
$\tan \beta = v_2/v_1$. 
\par
The stationary condition of the phase 
\begin{equation}
\label{stationaryconditionp}
 \left.{\partial V \over \partial \phi}\right| = 0  
\end{equation}
induces 
\begin{equation}
\label{theta}
 \sin \phi = 0  \;\; {\rm or}\;\;
 \cos \phi = {2 m_{12}^2 - \lambda_6 v_1^2 - 
                       \lambda_7 v_2^2 \over 
                       4 \lambda_5 v_1 v_2}.
\end{equation}
The solution which has non-vanishing phase is 
derived from the second equation 
of Eq.(\ref{theta}) 
and we denote $\phi = \phi_0$ 
{}for this case. 
The necessary condition for 
spontaneous $CP$ violation is 
\begin{equation}
 \left. \langle V \rangle \right|^{\langle H_1 \rangle = v_1}_{
 \langle H_2 \rangle = v_2 (-v_2)}\; > \; \left. \langle V \rangle 
 \right|^{\langle H_1 \rangle = v_1}_{
 \langle H_2 \rangle = v_2 e^{i \phi_0}} 
\end{equation}
{}for $\phi_0 \neq 0, \pi$, which derives 
\begin{equation}
\label{Ncondition}
\lambda_5 > 0, \;\;\;\;\;\; 
\left| {2 m_{12}^2 - \lambda_6 v_1^2 - \lambda_7 v_2^2 \over 
                       4 \lambda_5 v_1 v_2} \right| < 1.
\end{equation}
It means that $m_{12}^2$ must be small of 
$O(\lambda_{5} v^2)$ in order 
to get spontaneous $CP$ violation. 
Eq.(\ref{Ncondition}) is just a necessary and 
not the sufficient condition for 
spontaneous $CP$ violation. 
We must not forget that there exist another stationary point 
with vanishing phase corresponding to the first 
equation of Eq.(\ref{theta}). 
\par
By the use of stationary conditions 
of VEVs 
\begin{equation}
\label{stationarycondition}
  \left.{\partial V \over \partial v_i}\right| = 0 \quad (i=1,2) , \\
\end{equation}
we eliminate $m_1^2$ and $m_2^2$ as 
\begin{eqnarray}
\label{m12}
 m_1^2 &=& {\overline{g^2} \over 2}(v_2^2 - v_1^2) 
         + \lambda_5 v_2^2 - 
        \left({3 \lambda_6 v_1 v_2 \over 2} + 
              {\lambda_7 v_2^3 \over 2 v_1} \right) 
        \cos \phi_0 , \\
 m_2^2 &=& {\overline{g^2} \over 2}(v_1^2 - v_2^2)
         + \lambda_5 v_1^2 - 
        \left({\lambda_6 v_1^3 \over 2 v_2} + 
              {3 \lambda_7 v_1 v_2 \over 2} \right) 
        \cos \phi_0 , 
\end{eqnarray}
where $\overline{g}^2 \equiv (g^2+g'^2)/2$. 
Now we can decide specific Higgs potential with 
definite values of $m_1^2, m_2^2$, and 
$m_{12}^2 = 2 \lambda_5 v_1 v_2 \cos \phi_0 + 
(\lambda_6 v_1^2 + \lambda_7 v_2^2)/2$, 
which should have the stationary 
point at the non-trivial phase $\phi = \phi_0$. 
Expanding fields around this point as 
Eq.(\ref{fielddifinition0}), 
mass spectra become 
\begin{eqnarray}
\label{mat2}
 M_{S1-S1}^2 &=& \overline{g^2} v_1^2 + 2 (
       \lambda_5 v_2^2 \cos^2 \phi_0 + 
       \lambda_6 v_1 v_2 \cos \phi_0), \\
 M_{S2-S2}^2 &=& (\overline{g^2} + \Delta) v_2^2 + 2 (
       \lambda_5 v_1^2 \cos^2 \phi_0 + 
       \lambda_7 v_1 v_2 \cos \phi_0), \\
 M_{S1-S2}^2 &=& - {\overline{g^2}\over 2} v_1 v_2 - 
       2 \lambda_5 v_1 v_2 \sin^2 \phi_0 + 
       \lambda_6 v_1^2 \cos \phi_0 + 
       \lambda_7 v_2^2 \cos \phi_0,      \\
 M_{S1-A}^2  &=& -(2 \lambda_5 \cos \phi_0 v_2 + 
                     \lambda_6 v_1) 
                 v \sin \phi_0  ,              \\
 M_{S2-A}^2  &=& -(2 \lambda_5 \cos \phi_0 v_1 + 
                     \lambda_7 v_2) 
                 v \sin \phi_0  ,              \\
\label{mat222}
 M_{A-A}^2   &=& 2 \lambda_5 v^2 \sin^2 \phi_0 . 
\end{eqnarray}
$\Delta$ represents the top and stop effects 
\begin{equation}
\label{DELTA} 
\Delta \equiv {3 h_t^4 \over 4 \pi^2} \; {\rm ln}\: 
              {m_t^2 + m_{\tilde t}{}^2 \over m_t^2} \;, 
\end{equation}
where $m_{\tilde t}$ is the soft breaking stop mass parameter. 
Eq.(\ref{DELTA}) is derived from the 
one loop effective potential\cite{COLMAN} 
including only top and stop contributions, that is 
\begin{equation}
\label{topeffect}
 V_{\rm top} = {3 \over 16 \pi^2} 
    \left[ (h_t^2 |H_2|^2 + m_{\tilde t}^2)^2 {\rm ln} 
          {(h_t^2 |H_2|^2 + m_{\tilde t}^2) \over Q^2} - 
            h_t^4 |H_2|^4 {\rm ln} 
          {h_t^2 |H_2|^2 \over Q^2} \right] \ ,
\end{equation}
where stop left-right mixing are neglected. 
The values of $M_{S1-A}^2$, $M_{S2-A}^2$ and $M_{A-A}^2$ are 
same as calculated by Pomarol\cite{POMAROL}. 
\par
Next we show that 
if top and stop radiative corrections 
are not included, 
$CP$ violating vacuum with non-vanishing phase can not be 
a global minimum. 
We expand the determinant by small parameters 
$\lambda_{5 \sim 7}$ as 
\begin{equation}
\label{detsp}
 {\rm Det} M_{ij}^2 = {\rm Det^{(0)}} M_{ij}^2 + 
                      {\rm Det^{(1)}} M_{ij}^2 + 
                      {\rm Det^{(2)}} M_{ij}^2 + ..... \;\; . 
\end{equation}
As for the order $O(\lambda_{5 \sim 7}^0)$, 
${\rm Det^{(0)}} M_{ij}^2 =0$. 
It is the result from so-called Georgi-Pais theorem\cite{GP}, 
which says that the radiative symmetry breaking can be possible 
only when massless particle exists in the tree level. 
As for $O(\lambda_{5 \sim 7}^1)$, 
\begin{equation}
\label{detsp1}
 {\rm Det^{(1)}} M_{ij}^2 = 
      2 \lambda_5 \overline{g^2} \Delta
      v^2 v_1^2 v_2^2 \sin^2 \phi_0 .
\end{equation}
And for the next order $O(\lambda_{5 \sim 7}^2)$, 
\begin{equation}
\label{detsp2}
 {\rm Det^{(2)}}M_{ij}^2 = 
      - \overline{g^2}
      [ 8 \lambda_5^2 + (\lambda_6 + \lambda_7)^2 ] 
      v^2 v_1^2 v_2^2 \sin^2 \phi_0 .
\end{equation}
${\rm Det^{(1)}} M_{ij}^2$ is positive definite and 
${\rm Det^{(2)}} M_{ij}^2$ is negative definite. 
In order for $CP$ violating vacuum to be stable, 
where is a global minimum in fact, 
the relation 
\begin{equation}
 {\rm Det^{(1)}} M_{ij}^2 > |{\rm Det^{(2)}} M_{ij}^2| 
\end{equation}
must be satisfied. 
{}For this inequality to be satisfied, 
top and stop contributions are essential, and 
stop mass must be larger than 178 GeV at $\tan \beta = 1$ 
(188 GeV at $\tan \beta = \infty$) when $m_t = 174$ GeV. 
Otherwise the determinant of this neutral Higgs mass 
matrix becomes negative. 
Without top and stop contributions, 
the stationary point which break $CP$ symmetry 
is not the true vacuum and $CP$ conserving point 
corresponding to 
the first equation of Eq.(\ref{theta}) 
becomes the true vacuum. 
{}For example, in the case of $\phi_0 = \pi/2$, 
we can really show 
\begin{eqnarray}
 & & \left. \langle V \rangle \right|^{
     \langle H_1 \rangle = v_1}_{
     \langle H_2 \rangle = v_2 e^{i \phi_0}} - 
     \left. \langle V \rangle \right|^{
     \langle H_1 \rangle = v_1^2 - v_2^2}_{ 
     \langle H_2 \rangle = 0} 
            = \lambda_5 v_2^4 + O(\lambda_5^2) \; > 0   
             \;\;\;\;  (v_1^2 > v_2^2) , \\
 & & \left. \langle V \rangle \right|^{
     \langle H_1 \rangle = v_1}_{
     \langle H_2 \rangle = v_2 e^{i \phi_0}} - 
     \left. \langle V \rangle \right|^{
     \langle H_1 \rangle = 0}_{
     \langle H_2 \rangle = v_2^2 - v_1^2} 
                = \lambda_5 v_1^4 + O(\lambda_5^2) \; > 0   
             \;\;\;\;  (v_2^2 > v_1^2) ,
\end{eqnarray}
where we neglect $\lambda_{6,7}$ for simplicity. 
We stress that spontaneous 
$CP$ violation can not occur only by 
one loop diagram of the chargino and the neutralino 
contrary to Refs.\cite{MAEKAWA}\cite{POMAROL}. 
Top and stop effects are essentially needed. 
However these effects do not influence 
to the $M_{A-A}^2$ component at all. 
Then the lightest Higgs mass has little 
dependence of $m_{\tilde t}$, and 
its mass becomes 
smaller than about 5.5 GeV. 
\par
There is no allowed region in 
$\tan \beta \geq 1$ which satisfies 
following experimental constraints; 
(A): the branching ratio 
$B(Z \rightarrow h_1h_2)$ should be less than 
$10^{-7}$\cite{PDG}, 
(B): $B(Z \rightarrow h_i l^+l^-)$ should be smaller than 
$1.3 \times 10^{-7}$\cite{PDG}\cite{ALEPH}, 
where $h_1$ and $h_2$ 
are lightest and second lightest 
physical Higgs states, respectively. 
However in $\tan \beta < 1$, 
there is allowed region, 
{}for example, 
\begin{equation}
\label{agree}
 \tan \beta = 0.2, \;\;\; \phi_0 = \pi/2, 
 \;\;\;m_{\tilde t} = 3 \:{\rm TeV} . 
\end{equation}
But in this case, $h_t/4 \pi^2 \simeq 1.35$, 
so we can not trust the loop expansion 
of Eq.(\ref{topeffect}). 
\par
How does the situation change if 
the stop left-right mixing is included? 
Are there possibilities that there appears 
experimentally allowed region in $\tan \beta \geq 1$ 
by additional parameters 
$A_t$ and $\mu$ appeared 
in Eqs.(\ref{mat2})$\sim$(\ref{mat222})? 
Here $A_t$ is the SUSY breaking parameter of 
stop-stop-Higgs interaction. 
In order for $A_t$ and/or $\mu$ to have large 
effects on Higgs masses, 
they must be of 
$O(m_{\tilde t})$, 
since stop left-right mixing is 
proportional to $m_t \: A_t$ and $m_t \: \mu$. 
However it is shown that 
$CP$ violating vacuum becomes 
unstable in the parameter region of 
$A_t /m_{\tilde t}$ and/or 
$\mu /m_{\tilde t} \geq O(1/3)$ from the numerical analysis. 
And in the region of 
$A_t /m_{\tilde t}$ and $\mu /m_{\tilde t} \leq O(1/3)$, 
the situation is almost same as 
the limit case of small left-right mixing. 
In addition, 
the magnitude of $\lambda_5$, which is proportional to 
$M_{A-A}^2$, itself becomes 
small if stop left-right mixing exists. 
Thus, there is no experimentally allowed region 
even if parameters $A_t$ and $\mu$ take 
any values. 
Therefore we can conclude that 
spontaneous $CP$ violation in the MSSM 
is excluded from experimental constraints.

%
%
\section{Summary and Discussion}

We show that $CP$ violating vacuum can not 
be the true vacuum only by the chargino and the neutralino 
contributions. 
Top and stop contributions are crucially 
needed for spontaneous $CP$ violation 
in the MSSM. 
In the limit of small stop left-right mixing, 
the stop mass must be larger than about 
180 GeV for the vacuum stability, 
however, there is no experimentally allowed region 
in $\tan \beta \geq 1$. 
If we include the stop left-right mixing, 
additional parameters $A_t$ and $\mu$ 
appear in Higgs masses. 
However numerical analysis shows that 
both $A_t$ and $\mu$ should be smaller than $O(m_{\tilde t}/3)$ 
{}for the vacuum stability, 
and the situation 
is not so changed as the limit case of small stop 
left-right mixing. 
Thus, there is no experimental allowed region 
for spontaneous $CP$ violation in the MSSM 
in $\tan \beta \geq 1$. 
In order to obtain experimentally consistent 
spontaneous $CP$ 
violation scenario in the SUSY model, 
we should extend the MSSM to, for example, 
the next-to-minimal 
supersymmetric standard 
model (NMSSM)\cite{NMSSM} 
which contains an additional 
gauge singlet field $N$. 
In Refs.\cite{SPNMSSM}\cite{SPNMSSM2}\cite{SPNMSSM3}, 
they discuss spontaneous $CP$ violation 
in the NMSSM. 
Especially for the NMSSM with the 
scale invariant superpotential\cite{SPNMSSM2}\cite{SPNMSSM3}, 
spontaneous $CP$ violation 
occurs radiatively, 
so we can not avoid 
Georgi-Pais theorem. 
However the large VEV of $N$ can lift up the lightest 
Higgs mass, which is relatively 
light compared to $\langle N \rangle$ in actual, 
and spontaneous $CP$ violation in the 
NMSSM can be consistent with the experimental 
constraints (A) and (B)\cite{SPNMSSM3}. 
\par
As for explicit $CP$ violation 
in the Higgs sector of the MSSM, 
the mixing with scalars and a pseudoscalar 
appears also at the loop level. 
In this case $m_{12}^2$ does not need to be 
small of $O(\lambda_{5} v^2)$ as spontaneous $CP$ 
violation scenario. 
Angles of $CP$ mixings are negligibly small 
of $O(\lambda_{5 \sim 7})$\cite{POMAROL} 
as shown in Appendix, 
which are too small to influence the phenomenology

\vskip 1 cm
\noindent
{\bf Acknowledgements}\par
I would like to thank Professor A. I. Sanda 
for useful discussions and careful reading of manuscripts. 

%
\appendix\section{Explicit $CP$ Violation}

Once $CP$ symmetry is violated explicitly 
at the Lagrangian level, 
Eq.(\ref{lambdarelation}) is not held in general. 
It is because $\lambda_{5 \sim 7}$ 
are derived from 
various different diagrams containing 
different $CP$ phases such as Yukawa couplings, 
A terms, gaugino masses, and so on. 
We denote complex parameters $m_{12}^2$ and 
$\lambda_{5 \sim 7}$ as 
\begin{equation}
\label{lambda567ex}
m_{12}^2 \equiv   m_{12}^2 e^{i \varphi_{m}} , \;\; 
\lambda_5 \equiv \lambda_5 e^{i \varphi_5} , \;\; 
\lambda_6 \equiv \lambda_6 e^{i \varphi_6} , \;\; 
\lambda_7 \equiv \lambda_7 e^{i \varphi_7} , 
\end{equation}
where $m_{12}^2$ and $\lambda_{5 \sim 7}$ 
on the right hand side 
are real and positive parameters. 
By the phase rotation of Higgs fields, 
we can always take 
$\varphi_{m} = 0$. 
The stationary condition of the phase 
Eq.(\ref{stationaryconditionp}) 
induces 
\begin{equation}
\label{Pcondition}
 2 m_{12}^2 \sin \phi - 2 \lambda_5 v_1 v_2 
           \sin (\phi + 2 \varphi_5) - 
 \lambda_6 v_1^2 \sin (\phi + \varphi_6) - 
 \lambda_7 v_2^2 \sin (\phi + \varphi_7) = 0 . 
\end{equation}
It means that $\phi = 0$ can be the minimum, 
and which we take here for simplicity. 
Using Eq.(\ref{stationarycondition}) and (\ref{Pcondition}), 
the mass matrix of the neutral Higgs becomes 
\begin{eqnarray}
\label{mat2ex}
 M_{S_1-S_1}^2 &=& \overline{g^2} v_1^2 + 
       m_{12}^2 {v_2 \over v_1} + 
       {3\over2} \lambda_6 v_1 v_2 \cos \varphi_6 -
       {1\over2} \lambda_7{v_2^3 \over v_1} \cos \varphi_7  , \\
 M_{S_2-S_2}^2 &=& (\overline{g^2} + \Delta) v_2^2 + 
       m_{12}^2 {v_1 \over v_2} - 
       {1\over2} \lambda_6{v_1^3 \over v_2} \cos \varphi_6 + 
       {3\over2} \lambda_7 v_1 v_2 \cos \varphi_7 , \\
 M_{S_1-S_2}^2 &=& - m_{12}^2 - (\overline{g^2} - 
       {\lambda_5 \over 2} \cos \varphi_5) v_1 v_2 + 
       {3\over2}(\lambda_6 v_1^2 \cos \varphi_6 + 
                 \lambda_7 v_2^2 \cos \varphi_7 ) ,      \\
 M_{S_1-A}^2  &=& -{v \over 2 v_1} 
       (\lambda_6 v_1^2 \sin \varphi_6 - 
        \lambda_7 v_2^2 \sin \varphi_7) , \\
 M_{S_2-A}^2  &=& {v \over 2 v_2} 
       (\lambda_6 v_1^2 \sin \varphi_6 - 
        \lambda_7 v_2^2 \sin \varphi_7) , \\
\label{MAAinMSSM}
 M_{A-A}^2    &=& {v^2 \over 2 v_1 v_2}
       (2 m_{12}^2 - 4 \lambda_5 v_1 v_2 \cos \varphi_5 - 
       \lambda_6 v_1^2 \cos \varphi_6 - 
       \lambda_7 v_2^2 \cos \varphi_7) . 
\end{eqnarray}
The physical charged Higgs field is defined as 
$C^+ \equiv \cos \beta H^+ + \sin \beta H^{-*}$ and 
its mass is given by 
\begin{equation}
  m_C^2 = m_W^2 + M_{A-A}^2 .
\end{equation}
Mixings of scalars and a pseudoscalar are 
small of order $O(\lambda_{5 \sim 7} v^2)$. 
It is because $CP$ violation in the Higgs 
sector is not realized till 
radiative corrections are included. 
Contrary to spontaneous $CP$ violation, 
the determinant of 
$O(\lambda_{5 \sim 7}^0)$ becomes positive as 
\begin{equation}
\label{detex}
{\rm Det^{(0)}} M_{ij}^2 = {\overline{g^2} m_{12}^4 v^2 
                            (v_1^2 - v_2^2)^2 \over 
                             v_1^2 v_2^2} \;\; > \; 0 .
\end{equation}
Angles of $CP$ mixings in the Higgs sector are 
of $O(\lambda_{5 \sim 7}^0)$ from Eqs.(\ref{mat2ex})
$\sim$(\ref{MAAinMSSM}). 
Then the effect of $CP$ violation in the Higgs 
sector is negligibly small comparing to other sectors.

%

\end{document}